\documentclass{PoS}
\usepackage{amsmath}

\title{The sphaleron rate at the electroweak crossover with 125\,GeV Higgs mass}

\ShortTitle{The sphaleron rate at the electroweak crossover with 125\,GeV Higgs mass}

\author{\speaker{Michela D'Onofrio}\\
        Department of Physics, University of Helsinki and Helsinki Institute of Physics\\
        P.O. Box 64, FI-00014 Helsinki\\
        E-mail: \email{michela.donofrio@helsinki.fi}}

\author{Kari Rummukainen\\
		Department of Physics, University of Helsinki and Helsinki Institute of Physics\\
        P.O. Box 64, FI-00014 Helsinki\\        
        E-mail: \email{kari.rummukainen@helsinki.fi}}

\author{Anders Tranberg\\
        Niels Bohr Institute, University of Copenhagen\\
        Blegdamsvej 17, DK-2100 Copenhagen\\
        and Helsinki Institute of Physics\\
        P.O. Box 64, FI-00014 Helsinki\\ 
        E-mail: \email{antranbe@nbi.dk}}

\abstract{We measure the sphaleron rate with the physical parameters of the Standard Model. In particular, we plug into the calculations the recently found Higgs mass $m_H =$ 125\,GeV. The sphaleron rate tells us about the efficiency of baryon number violation through sphaleron transitions. These occur above the electroweak scale $\sim$ 100\,GeV and get exponentially suppressed at temperatures substantially below the electroweak crossover. The sphaleron rate enters computations of Baryogenesis via Leptogenesis, where it converts non-zero lepton number into non-zero baryon number. We simulate the effective electroweak theory on the lattice with multicanonical and real-time methods to calculate the sphaleron rate as a function of temperature through the electroweak crossover.}

\FullConference{The 30th International Symposium on Lattice Field Theory\\
                 June 24 - 29,  2012\\
                 Cairns, Australia}

\begin{document}

\section{Introduction}
Baryon and lepton numbers are not exactly conserved quantities in the Standard Model, because of the axial anomaly, which connects them to the Chern-Simons number of the weak gauge field. Vacua in the electroweak theory are labeled by an integer-valued Chern-Simons number
\begin{equation}
N_{CS} = \int d^3 x \ j_{CS}^0 = -\frac{g^2}{64 \pi}\int d^3x \ \epsilon^{ijk} \textrm{Tr}\left(A_i F_{jk} + i \frac{g}{3} A_i A_j A_k \right).
\end{equation}
In the early Universe, for $T >$ 100\,GeV, it is possible to move from one vacuum to another surmounting the potential barrier through sphaleron transitions, thus changing Chern-Simons number, B and L according to
\begin{equation}
\frac{1}{n_G}\left[B(t)-B(0)\right]=L_i(t)-L_i(0)
=N_{\rm CS}(t)-N_{\rm CS}(0),
\label{eq:ncs}
\end{equation}
with $n_G$ the number of generations of fermions. At zero temperature, where the potential barrier is high, the rate of Chern-Simons number fluctuations, the sphaleron rate, is exponentially suppressed and, when the Higgs field expectation value $v \gg T$, the rate is negligible. In electroweak Baryogenesis scenarios \cite{Kuzmin:1985mm} the baryon number of the Universe is generated during the electroweak phase transition.  However, this scenario does not work in the Standard Model: it requires stronger CP violation and first order phase transition, whereas the Standard Model has a smooth crossover \cite{Kajantie:1996mn}.

Nevertheless, the sphaleron rate during the electroweak crossover in the Standard Model is 
relevant for Baryogenesis via Leptogenesis: indeed the sphaleron rate converts lepton asymmetry into baryon asymmetry.  If the lepton asymmetry is generated just before or during 
the electroweak phase transition, how the sphaleron rate shuts off has an effect on the generated baryon number.

The sphaleron rate has been studied in the broken phase before, but either with unphysical Higgs masses \cite{Moore:1998swa,Moore:2000jw,Tang:1996qx} or not very deeply in the broken phase \cite{Moore:1998swa}. Both perturbative calculations \cite{Burnier:2005hp,Philipsen:1995sg} and lattice simulations \cite{Moore:1998swa,Ambjorn:1990pu,Krasnitz:1993mt,Shanahan:1998gj} have been used.

In this paper we follow the procedure of our previous works \cite{D'Onofrio:2010es,D'Onofrio:2012jk} in determining the sphaleron rate at the energy range of the electroweak crossover, with the crucial difference that now we are able to perform the simulations with the recently found Higgs mass, $m_H =$ 125\,GeV. Our results are compared to analytical estimates both in the broken and symmetric phases \cite{Burnier:2005hp}.

\section{Theory on the lattice}
The thermodynamics of the 4-dimensional electroweak theory is studied in 3 dimensions by means of dimensional reduction \cite{Kajantie:1995dw}, a perturbative technique that gives the correspondence between 4D and 3D parameters. The result is a SU(2) effective theory with the Higgs field $\phi$ and gauge field $A_{\mu}$ ($F_{ij}$)
\begin{equation}
L = \frac{1}{4} F^a_{ij} F^a_{ij} + (D_i\phi)^{\dagger}(D_i\phi)+m_3^2 \phi^\dagger \phi + \lambda_3 (\phi^\dagger \phi)^2,
\end{equation}
and 3D effective parameters $g_3^2$, $\lambda_3$ and $m_3^2$. 

B\"odeker showed \cite{Bodeker:1998hm} that at leading order in log(1/$g$) the time evolution of this effective SU(2) Higgs model is governed by Langevin dynamics. The latter, though, is very slow on the lattice and can be substituted by any other dissipative procedure, e.~g.~heat bath. 
One heat-bath sweep through the lattice corresponds to the real-time step \cite{Moore:2000jw}
\begin{equation}\label{Moorep28}
\Delta t = \frac{a^2 \ \sigma_{el}}{4}, 
\end{equation}
with
\begin{equation}\label{Moore3.9}
\sigma^{-1}_{\textrm{el}} = \frac{3}{m^2_D}\gamma , \quad  \quad \gamma= \frac{N g^2 T}{4 \pi} \left[ln\frac{m_D}{\gamma}+3.041 \right]
\end{equation}
where $\sigma_{\textrm{el}}$ is the non-abelian color conductivity, which quantifies the current response to infrared external fields, $N$ is the dimension of the SU(N) gauge group, and $m_D$ is the Debye mass, determining the length scale $l_D$ $\sim$ $1/m_D$ $\sim$ $1/gT$. We made use of a 32$^3$ lattice, with $\beta_G \equiv \frac{4}{g_3^2 a} =$ 9, where $g_3$ is the 3D gauge coupling and $a$ the lattice spacing. In real-time simulations, for each temperature we computed 4 trajectories for every 1000 initial configurations.

\section{Measurement of the sphaleron rate}
In the symmetric phase we make use of canonical Monte Carlo simulations and approach the broken phase. 
At very low temperatures, the rate is highly suppressed and canonical methods do not work anymore. 
We need multicanonical methods, which calculate a weight function that compensates the high potential barrier between the vacua, thus allowing transitions.
The exact value of the sphaleron rate 
\begin{equation}
\Gamma \equiv \lim_{t\rightarrow \infty} \frac{\langle(N_{CS}(t)-N_{CS}(0))^2\rangle}{V \ t}
\end{equation}
is obtained, in the broken phase, through a method similar to the one used in \cite{Moore:1998swa,Moore:2000jw}. 

\begin{enumerate}
\item Once done the multicanonical
  simulations, we obtain the canonical (physical) probability distribution
  of the Chern-Simons number \ $p_{\rm phys.}(N_{\rm CS})$. 

\item We choose a narrow interval $1/2 - \epsilon/2 \le N_{\rm CS} \le
  1/2 + \epsilon/2$ around the point that separates vacuum $N_{\rm CS}
  = 0$ from the vacuum $N_{\rm CS}=1$. The relative probability of
  finding a configuration here is
  \begin{equation}
    P(|N_{\rm CS} - 1/2| < \epsilon/2) =
    \int_{1/2-\epsilon/2}^{1/2+\epsilon/2} dN p_{\rm phys}(N).
  \end{equation}
  This is where we need multicanonical
  methods, as the probability of being on top of the
  barrier is extremely small, and to get a reliable estimate would
  take an impractically long time with canonical sampling. 

\item Let us now take a random configuration from the canonical
  distribution but with the constraint $1/2 - \epsilon/2 < N_{\rm CS}
  < 1/2 + \epsilon/2$; i.e.~near the top of the potential barrier. Starting from this
  configuration, we now generate two real-time trajectories using heat-bath dynamics.
  The trajectories are evolved until the
  Chern-Simons number falls near a vacuum value.  Interpreting one of
  the trajectories as evolving backwards in time, we can glue the
  trajectories together at the starting point and obtain a
  vacuum-to-vacuum trajectory.  The trajectory can either return to
  the starting vacuum or be a genuine tunneling trajectory. Only the latter-type trajectories contribute to the
  sphaleron rate.

\item We can obtain the tunneling rate by measuring 
  $|\Delta N_{\rm CS} / \Delta t|$ from the trajectories at the moment they cross the
  value $N_{\rm CS}=1/2$.  Here $\Delta t$ is the time interval between successive measurements, and $\Delta N_{\rm CS}$ the change in Chern-Simons number.  This characterizes the probability flux throughout 
  the top of the barrier.   We obtain the physical time difference from
  the relation between the heat-bath ``time'' and physical time, equation~(\ref{Moorep28}).
  
\item
  If the tunneling trajectories would go straight across the top, the ingredients above would be sufficient to calculate the total rate.  However, typically the trajectories ``random walk'' near the top of the barrier and can cross the value $N_{\rm CS}=1/2$ several times.  Because the trajectories were chosen starting from a set of configurations near the top of the barrier, this leads to overcounting: the evolution could be started at any point the $N_{\rm CS}=1/2$ limit is crossed.  This can be compensated by calculating a dynamical prefactor 
  \begin{equation}\label{dynpref}
  \textrm{d} = \frac{1}{N_{\rm traj}} \sum_{\rm traj} \frac{\delta_{\rm tunnel}}{\# \ \textrm{crossings}},
\end{equation}
where the sum goes over the ensemble of trajectories, $N_{\rm traj}$ is the number of trajectories, $\delta_{\rm tunnel}$ is 0 if the trajectory does not lead to a change of the vacuum and 1 if it does, and (\#\,crossings) is the number of times the trajectory crosses $N_{\rm CS}=1/2$.  
\end{enumerate}
With these ingredients, the sphaleron rate now becomes
\begin{equation}\label{Gammamc}
\Gamma = 
\frac{P(|N_{\rm CS} - 1/2| < \epsilon/2) }{\epsilon}  
\left\langle  \left| \frac{\Delta N_{\rm CS}}{\Delta t} \right| \right\rangle  \textrm{d}.
\end{equation}
We note that the result is independent of $\epsilon$ as long as $\epsilon \ll 1$.
It is also independent of the frequency $\Delta t$ with which the Chern-Simons number is measured:  if we decrease the measurement interval, the trajectories become more jagged due to the random-walk nature of the heat-bath updates.  This will increase the number of the crossings of the value $N_{\rm CS}~=~1/2$ and hence decrease $\textrm{d}$.  However, the latter is completely compensated by a corresponding increase in $\langle | \Delta N_{\rm CS}/\Delta t|\rangle$.  If the measurement interval $\Delta t$ is small enough, random walk arguments imply $\textrm{d} \propto (\Delta t)^{1/2}$ and $\langle | \Delta N_{\rm CS}/\Delta t|\rangle \propto (\Delta t)^{-1/2}$.  This is corroborated by the numerical data.  Thus, equation~(\ref{Gammamc}) has a well-defined continuum limit.

\begin{figure}
  \begin{center}
  \includegraphics[width=.8\textwidth,viewport=3 0 490 460,clip]{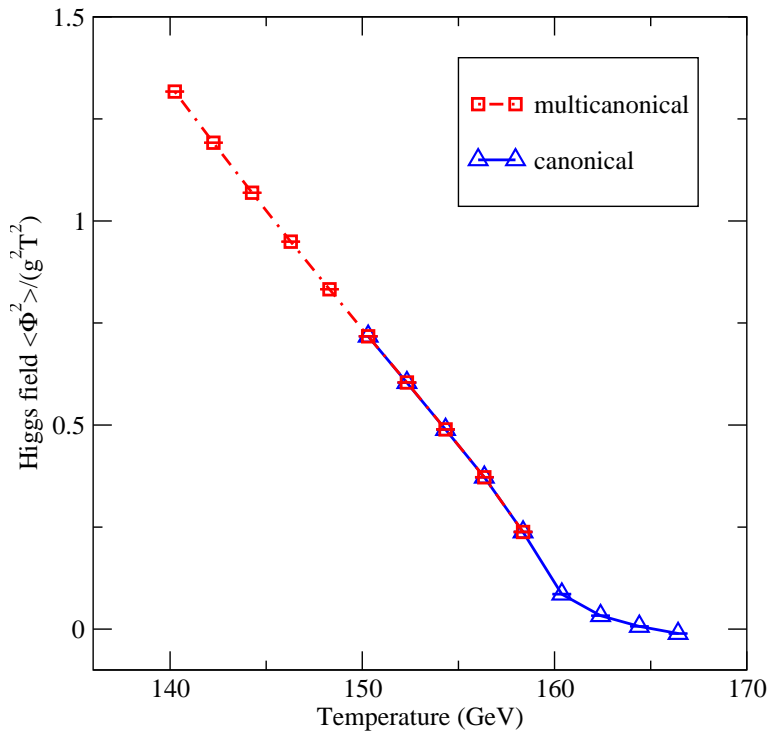}
  \end{center}
\caption{The Higgs expectation value $\langle\phi^2\rangle$ for Higgs mass of 125 GeV as a function of temperature. The high-temperature canonical and low-temperature multicanonical results match beautifully in the transition region.}
\label{fig:m125phi}
\end{figure}

\section{Results}
\begin{figure}
  \begin{center}
  \includegraphics[width=.8\textwidth,viewport=3 0 490 460,clip]{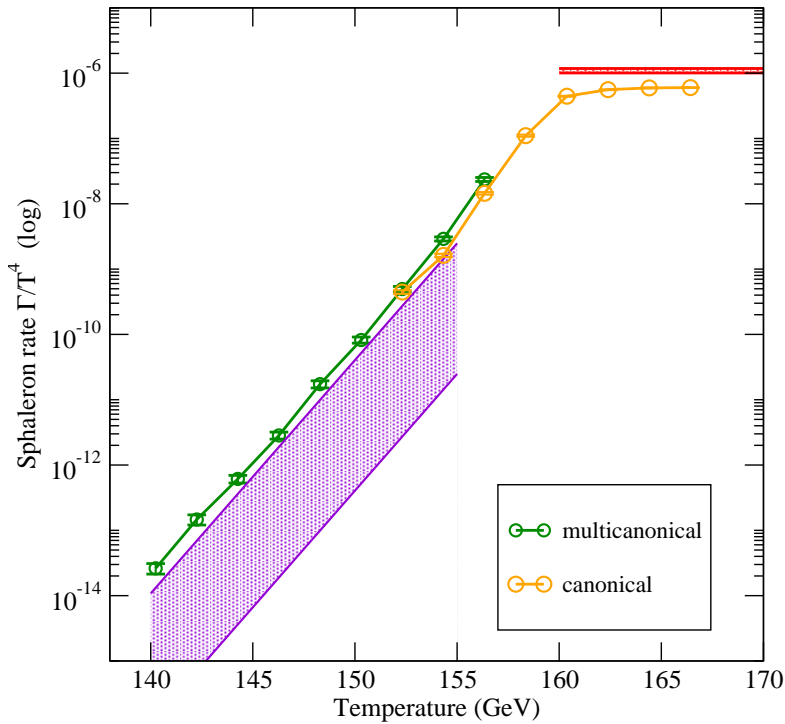}
  \end{center}
  \caption{The sphaleron rate for a Higgs mass of 125 GeV. The high-temperature canonical and low-temperature multicanonical results again match very well in the transition region. Also shown are previous high-temperature estimates $\sim$ 5 $\times$ 10$^{-7}$ T$^4$, (top, horizontal line) and perturbative calculations in the low-temperature phase (bottom, wide band), both from \cite{Burnier:2005hp}.}
  \label{fig:m125rate}
\end{figure}

We obtain the sphaleron rate $\Gamma / T^4$ and the expectation value for the Higgs field $\langle\phi^2\rangle$ for Higgs mass $m_H =$ 125\,GeV. Figure \ref{fig:m125phi} shows the Higgs field expectation value behavior as a function of temperature. We start from the ``symmetric phase'' with canonical Monte Carlo simulations and lower the temperature to reach the ``broken phase'', where we switch to multicanonical simulations. We can see the Higgs field assuming a non-zero value when approaching the broken phase, and the transition from the two methods occurring smoothly. \\
The sphaleron rate as a function of temperature is shown in Figure \ref{fig:m125rate}. Here again we perform the simulations with canonical Monte Carlo at high temperatures and continue with multicanonical methods when reaching the cold broken phase. The sphaleron rate changes from its asymptotic value to become exponentially suppressed at very low temperatures. The canonical and multicanonical methods are in good agreement. 
The theoretical curves, used to compared our results, were obtained separately for the broken and symmetric phases, through perturbative calculations in \cite{Burnier:2005hp}. 

\section{Conclusion}
We improved the previous estimates for the sphaleron rate and determined its behavior from the symmetric to the broken phase, through the electroweak crossover. Our results are in agreement with previous estimates in the symmetric phase, and in the broken phase the slope of our curve is the same as in the analytic calculation \cite{Burnier:2005hp}. 

Even though the Standard Model has a too weak source of CP-violation in the quark sector, Baryogenesis might still be viable through lepton number violating processes. The sphaleron rate plays an important role in Leptogenesis, as the conversion of lepton to baryon number depends on it, and it is therefore important to know its size rather accurately.

\acknowledgments This work is supported by the Academy of Finland
grants 114371 and 1134018.  The computations have been made at the
Finnish IT Center for Science (CSC), Espoo, Finland. M.\,D.~acknowledges support from the Magnus Ehrnrooth foundation. A.\,T.~is supported by the Carlsberg Foundation.

\end{document}